# Effect of chain length on fragility and thermodynamic scaling of the local segmental dynamics in poly(methylmethacrylate)

R. Casalini[1,2], C.M. Roland[2], and S. Capaccioli[3]

[1]*George Mason University, Chemistry Department, Fairfax VA  22030*
[2]*Naval Research Laboratory, Chemistry Division, Code 6120, Washington DC 20375-5342*
[3]*Dipartimento di Fisica, Università di Pisa, Pisa, Italy and CNR-INFM, CRS SOFT, Università di Roma "La Sapienza", Piazzale Aldo Moro, Roma, Italy.*

January 29, 2007

Abstract

Local segmental relaxation properties of poly(methylmethacrylate) (PMMA) of varying molecular weight are measured by dielectric spectroscopy, and analyzed in combination the equation of state obtained from PVT measurements. The usual variation of glass transition temperature and fragility with molecular weight are observed. We also find, in accord with the general properties of glass-forming materials, that a single molecular weight dependent scaling exponent, $\gamma$, is sufficient to define the mean segmental relaxation time, $\tau_\alpha$, and its distribution. This exponent can be connected to the Grüneisen parameter and related thermodynamic quantities, thus demonstrating the interrelationship between dynamics and thermodynamics in PMMA. Changes in the relaxation properties ("dynamic crossover") are observed as a function of both temperature and pressure, with $\tau_\alpha$ serving as the control parameter for the crossover. At longer $\tau_\alpha$ another change in the dynamics is apparent, associated with a decoupling of the local segmental process from ionic conductivity.

**Introduction**

The chain connectivity of polymers introduces complexities that make their behavior distinct from that of molecular liquids. For example, the uncrossability of long chains gives rise to entanglement constraints, which confer shear-thinning and marked viscoelasticity to the low frequency dynamics, effects generally absent in small molecules. On the other hand, the structural relaxation properties of polymers and molecular liquids are virtually indistinguishable. The dynamics of both materials exhibit a dramatic slowing down upon approach to the glass transition, which is unaccompanied by any marked change in structure or molecular configuration. A common structural relaxation property is the fragility, $m = \partial \log(\tau_\alpha) / \partial (T_g/T)\big|_{T=T_g}$, quantifying the temperature dependence of the structural relaxation time, $\tau_\alpha$.[1] Polymers tend to be somewhat more fragile (larger $m$) than molecular liquids, although there are exceptions.[2] The fragility of polymers can vary with their molecular weight ($M_w$). For example, $m$ increases with $M_w$ for polystyrene (PS),[3,4,5] poly(methymethacrylate) (PMMA),[6] polypropylene glycol (PPG),[7] and methyl-terminated PPG (which lacks H-bonds).[8] More flexible chain polymers, such as polydimethylsiloxane (PDMS)[9] and polyphenylmethylsiloxane (PMPS),[10] lack this sensitivity of $m$ to $M_w$. This is consistent with the observation that less flexible chains and those having bulky pendant groups tend to exhibit more fragile behavior.[11] This effect is ascribed to stronger intermolecular constraints for the latter, as effected also by crosslinking.[12]

Although fragility quantifies the temperature dependence of the relaxation time, since an isobaric temperature variation also changes the density, $m$ provides no information about whether the dynamics are thermally activated or governed more by the volume changes accompanying changes in $T$. In particular for van der Waals polymers, shorter chains appear to allow easier segmental rearrangements, so that volume exerts a stronger effect for decreasing $M_W$.[13] A method used to quantify the relative influence of temperature, $T$, and specific volume, $V$, on the structural dynamics is the ratio of the isochoric activation energy to the isobaric activation enthalpy, $\dfrac{E_V}{H_P} = \dfrac{\partial \log(\tau)/\partial 1/T\big|_V}{\partial \log(\tau)/\partial 1/T\big|_P}$, by convention evaluated at the glass transition.[14] This ratio varies between 0 and 1 for the

limiting cases of volume- and temperature-dominated dynamics, respectively. Compilations of $E_V/H_P$ for many materials have been published.[15,16,17,18] For hydrogen bonded materials $E_V/H_P$ is close to unity, while for van der Waals molecules this ratio is smaller ($0.38 \leq E_V/H_P \leq 0.6$); polymers exhibit intermediate values ($0.52 \leq E_V/H_P \leq 0.86$).

Very flexible chains, such as those with oxygen in the backbone, have low ratios; for example, $E_V/H_P = 0.56$ and 0.52, respectively, for PDMS and PMPS. As temperature is increased above $T_g$, $E_V/H_P$ decreases.[13] This leads to an exceptional low value of $E_V/H_P = 0.25$ for poly(2,6-dimethyl-1,4-phenylene oxide) (PPO),[19] which combines a flexible backbone and a unusually high glass transition temperature ($T_g$=461.75K).

An alternative approach to evaluating temperature and volume dependences is by thermodynamical scaling of relaxation times.[20] For each material there is a constant $\gamma$ such that

$$\log(\tau_\alpha) = \Im(TV^\gamma) \qquad (1)$$

This scaling behavior is observed very generally for non-associated organic glass-formers (the exceptions being materials with extensive hydrogen bonding).[21] Eq.(1) can be derived from consideration of the $T$- and $V$-dependences of the entropy.[22,23,24] It follows from the idea that the repulsive part of the intermolecular potential dominates the local liquid structure,[25,26] so that for local properties the potential can be approximated with a spherically symmetric, two-body interaction [27,28]

$$U(r) = \varepsilon \left(\frac{\sigma}{r}\right)^n - \frac{a}{r^2} \qquad (2)$$

where $\varepsilon$ and $\sigma$ are the characteristic energy and length scale of the system, $r$ is the intermolecular distance, and $n$ (=3$\gamma$). The mean-field parameter $a$ describes the long-range attractive potential, which can be taken as constant when considering local properties, such as structural relaxation in liquids or the local segmental relaxation of polymers. Recent simulations of the glass transition have employed this inverse power repulsive potential.[25,29] Molecular dynamic simulations of 1,4-polybutadiene, employing a potential given by the superposition of a 6-12 Lennard-Jones intermolecular potential and harmonic chain stretching and bending potentials, yields the scaling behavior (eq.(1)) with $\gamma=2.8$.[30,31]

These results imply that for low molecular weight polymers, the potential is much closer to that of a monomer, approaching the spherically symmetric form of eq.(2), while for higher molecular weight the contribution of the intramolecular forces, modeled with additional harmonic terms,[21] is more significant. The effect of the latter is to reduce the steepness of the effective potential, with consequently lower $\gamma$ for high molecular weight polymers.

There is a simple relation between the scaling exponent $\gamma$ and the activation enthalpy ratio[20,32]

$$\left.\frac{E_V}{H_P}\right|_{T_g} = \left(1+\gamma\alpha_P T_g\right)^{-1} \qquad (3)$$

where $\alpha_P$ $(=\partial \ln(V)/\partial T|_P)$ is the isobaric thermal expansion coefficient. Since the product $\alpha_P T_g$ is approximately constant (= 0.18 ± 0.2 [33,34]), $\gamma$ has an inverse relationship to $E_V/H_P$. Since $\gamma$ is constant and $\alpha_P T_g$ generally increases with $T$, the reduction in $E_V/H_P$ with $T$, referred to above, follows directly from eq.(3);[13] that is, volume becomes more important at temperatures above $T_g$. Since the ratio $E_V/H_P$ can be calculated directly from the pressure- and temperature-dependences of the volume,[35] using eq.(3) $\gamma$ can be obtained without making relaxation measurements (i.e., from just thermodynamic data).[13]

In this work we measured structural relaxation times as a function of temperature and pressure, along with the equation of state, for three oligomers of PMMA. From these data we investigate the effect of chain length on the thermodynamical scaling and thus on the structural relaxation properties $E_V/H_P$, $\gamma$, and fragility.

**Experimental**

The three hydrogen-terminated PMMA (from Polymer Standard Service) had degrees of polymerization (*n*) equal to 3, 7, and 10, with weight average molecular weights of $M_w$=302 g/mol (trimer), $M_w$=402 g/mol (tetramer), and $M_w$=1,040 g/mol (decamer). The lower Mw samples were monodisperse, while the decamer had a polydispersity = 1.3. All samples were used as received

Pressure-volume-temperature (PVT) measurements were carried out with a Gnomix instrument.[36] At room temperature solid samples were molded under vacuum into a cylinder, while liquid samples were injected directly into the cell. The temperature was raised at 0.5 K/min at various fixed pressures up to 200 MPa. The ambient density was measured by the buoyancy method for the decamer and tetramer and volumetrically for the trimer.

For the tetramer, ambient pressure Dielectric spectra (($10^{-2}$ to $10^6$ Hz)) were measured with a Novocontrol Alpha Analyzer using a parallel plate configuration. Temperature was controlled using the Novocontrol Quatro cryosystem (± 0.01 K stability). Dielectric measurements on the other PMMA samples were carried out with the Alpha Analyzer, as well as an IMASS time domain dielectric analyzer ($10^{-4}$ to $10^3$ Hz) and an HP16453A test fixture with a HP4291A impedance analyzer ($10^6$-$10^9$ Hz). For these measurements at f<$10^6$ Hz, a closed-cycle helium cryostat with a helium atmosphere was used for temperature control to within ± 0.02 K. Measurements above $10^6$ Hz employed a ESPEC SH-240 temperature chamber in a nitrogen atmosphere.

For dielectric measurements at elevated pressure, the sample was contained in a Harwood Engineering pressure vessel, with hydraulic pressure applied using a Enerpac pump in combination with a pressure intensifier (Harwood Engineering). Pressures were measured with a Sensotec tensometric transducer (resolution = 150 kPa). The sample assembly was contained in a Tenney Jr. temperature chamber, with control to within ± 0.1 K at the sample.

Differential scanning calorimetry (DSC) employed a TA Instruments Q100, using liquid nitrogen cooling. Samples were cooled from the liquid state to below $T_g$ at 10 C/min. The absolute value of the heat capacity was obtained using a synthetic sapphire for calibration.[37]

**Results**

*Dynamic crossover.* Figs.1 (a) and (b) show representative dielectric loss spectra for the trimer and tetramer respectively, measured at atmospheric pressure. Comparing to spectra in the literature for high molecular weight PMMA,[38,39] a clear difference is evident in the dielectric strength of secondary relaxation relative to that of the α−relaxation. For high molecular weight PMMA the intensity of the secondary relaxation is larger than that of the structural relaxation, while for the oligomers it is much smaller, whereby it can be clearly resolved only in the vicinity of the glass transition. The effect of molecular weight and pressure on the secondary relaxation of PMMA will be discussed in detail in a subsequent publication; the focus herein is the α-relaxation.

The breadth of the α−peaks in Figure 1 decreases with temperature. A structural relaxation time (the most probable relaxation time, $\tau_\alpha$) can be defined from the maximum of the loss peak, with the latter determined by fitting the peak to an Havriliak-Negami function.[40] Any contribution from the secondary relaxation for the trimer and tetramer was neglected. For the decamer the spectra were analyzed using a single function only at higher temperatures where the secondary and structural peaks cannot be distinguished.

In Fig. 2a are the temperature dependences of $\tau_\alpha$ for the three oligomers, with the variation of $T_g$ over this range of $M_w$ evident A function to describe the temperature dependence of the relaxation time at constant pressure is the Vogel-Fulcher-Tamman-Hesse (VFTH) equation [41,42]

$$\tau(T) = \tau_0 \exp\left(\frac{DT_0}{T - T_0}\right) \quad (4)$$

where $T_0$, the Vogel temperature, $D$, and $\tau_0$ are constants. The validity of this equation is usually limited to temperatures close to the glass transition, since for shorter $\tau_\alpha$ (in the range $10^{-4}$ - $10^{-7}$ s) a second VFTH is required.[43,44,45,46,47] This deviation from eq.(4), referred to as the "dynamic crossover", is revealed using the derivative function

$$\phi_T = \left\{d\left[\log(\tau_\alpha)\right]/d\left[1000/T\right]\right\}^{-\frac{1}{2}} \quad (5)$$

introduced by Stickel et al.[45]. $\phi_T$ is a single straight line for data conforming to eq.(4). In Fig. 2b $\phi_T$ is displayed for the three oligomers. The trimer exhibits a change in dynamics at a temperature $T_B$ =264±5K corresponding to log($\tau_B$)=-6.9±0.3, as determined from the intersection of the two linear fits to the derivative data. For the tetramer a change in dynamics is less obvious; as seen in Fig. 2a, its relaxation times can be described by a single VFTH over the entire range (see inset). The Stickel plot for the tetramer gives a hint of a crossover at $T_B$ ~ 316K corresponding log($\tau_B$)~ -7.5, but the uncertainty is large.

For the decamer the range of the data is more limited and only a single VFTH is required to fit the $\tau_\alpha$. All parameters and values of $T_B$ are listed in Table 1.

*Ionic conductivity.* In the analysis of the loss spectra the dc-conductivity contribution from the presence of mobile ions in the polymer was included as an inverse power law, $\varepsilon_{dc}(\omega) = -i\sigma/\omega\varepsilon_0$, where $\varepsilon_0$ $(= 8.85419\ pF/m)$ is the vacuum permittivity and $\omega$ the angular frequency. In Fig. 3 this conductivity at atmospheric pressure is shown as a function of temperature. In the insert to Fig. 3 a plot of $\log(\sigma)$ *vs.* $\log(\tau_\alpha)$ shows that the expected inverse correlation of two quantities, with a linear coefficient close to -1 over most of the range. However, for the trimer and tetramer, some deviation is evident at $\tau_\alpha \sim 10^{-3}$s, with a smaller slope indicating that $\sigma$ is decreasing more slowly than $\tau_\alpha^{-1}$ as $T$ is reduced.

The usual interpretation of relationship between $\sigma$ and $\tau_\alpha$ is from consideration a simple hydrodynamic model for macroscopic (Brownian) particles. The product of the rotational relaxation time and the diffusion constant $D_T$ is constant according to the classical Stokes-Einstein (SE) and Debye-Stokes-Einstein (DSE) equations. These equations have been successfully applied to molecular motions in (low viscosity) liquids and to probes in a host of molecules having similar or smaller size.[48,49,50] The conductivity is proportional to the diffusion constant of the ions $D_i$ (Nernst–Einstein relation),[51]

$$D_i = \frac{\sigma kT}{ec} \qquad (6)$$

where $e$ and $c$ are the respective charge and concentration of the ions. For a supercooled liquid $D_i$ and $\sigma$ change by several orders of magnitude over a small range of $T$, while the change of $c$ is negligible (but not necessarily zero[52]); thus, $D_i\sigma \sim$ constant. Therefore, assuming $D_T \sim D_i$, then $\sigma\tau$ should be about constant.[53]

The behavior in the insert of figure 3 is consistent with $\sigma\tau \sim$ constant, while the deviation observed for the trimer and tetramer suggests a possible decoupling of translational and rotational motions for $\tau_\alpha > 10^{-3}$s, which is still far from the glass transition. A similar decoupling, manifested as an enhancement in the translational diffusion coefficient relative to the rate of rotational diffusion, has been reported at temperatures below $1.2T_g$ in various supercooled systems,[54,55] although recent numerical simulations suggest the need for a critical reexamination of this issue.[56]

*Glass transition temperature.* In Figure 4 is shown the variation of the glass transition temperature with $M_w$ as determined from calorimetry and dielectric measurements, using $\tau(T_g) = 100$s for the latter (for the decamer extrapolation of the fitted VFTH equation was required). Also included in the figure are literature results from DSC measurements of Andreozzi et al.[57]. There is good consistency among the different measurements, with $T_g$ exhibiting a strong dependence on chain length for $M_w < 10^4$ g/mol. Generally this dependence of $T_g$ on molecular weight is small for flexible chain polymers such as PDMS,[9] while more rigid or bulky polymers, for example polystyrene, have strongly $M_w$-dependent glass temperatures.[11] The usual interpretation is that chain ends confer extra unoccupied space ("free volume"), which has a large influence on $T_g$ for a rigid or bulky chain. This role of chain ends on $T_g$ is corroborated by studies of bidisperse polymer blends.[58] From free volumes ideas Fox and Flory[59] derived an expression for the variation of the $T_g$ of linear polymers with number average of molecular weight ($M_n$)

$$T_g = T_{g,\infty} - k_{FF} M_n \tag{7}$$

where $T_{g,\infty}$ is the limiting value of the glass transition temperature and $k_{FF}$ is a constant dependent on the chemical structure. A more accurate empirical expression was proposed by Ueberreiter and Kanig (UK)[60]

$$T_g^{-1} = T_{g,\infty}^{-1} + k_{UK} M_n \tag{8}$$

in which $k_{UK}$ is a constant.

In Fig.4 we show the fits of both equations, with the UK equation giving a better description of the data. The inset to fig.4 shows the change of the heat capacity at the glass transition $\Delta C_P$, versus $M_w$, including some literature data.[57,61,62,63] For low molecular weights ($<10^3$ g/mol), the heat capacity jump at the glass transition is constant within the limit experimental error, $\Delta C_P = 0.35 \pm 0.01$ Jg$^{-1}$C$^{-1}$. For higher $M_w$ (>1 kg/mol), $\Delta C_P$ decreases with molecular weight, in agreement with previous work, showing $\Delta C_P$ for PMMA decreasing from 0.37 to 0.28 J/gC for $M_w$ in the range from 1.45 to 55.9 kg/mol.[57] Results for other polymers are mixed: For PDMS [9] $\Delta C_P$ is independent of $M_W$, while for PS [3] $\Delta C_P$ decreases with increasing $M_W$.

The fragility, characterizing the temperature dependence of $\tau_\alpha$ close to $T_g$, was found to increase with molecular weight from 74 ± 2 for the trimer to substantially higher values for the high polymer, $m=115\pm16$ [61] and $m=145$.[64] A difference in fragility between low and high $M_w$ PMMA was preliminarily reported by Ding et al.[6]. The trend is similar to that observed for other polymers, as discussed in the introduction. As pointed out by Angell,[1,65] the fragility of liquids reflects the topology of the potential energy hypersurface governing the dynamics. This implies that more fragile liquids are associated with potential surfaces having a high density of minima, and hence a high configurational heat capacity change at $T_g$. However, the situation is complex, so that a simple correlation between $m$ and $\Delta Cp(T_g)$ is not realized.[66,67] Indeed we have shown for polymers that the two quantities are often anti-correlated for samples differing only in molecular weight.[9,64,68] Thus, the results herein confirm that the change of $m$ with $M_W$ is anti-correlated with the change of $\Delta C_P$ with $M_W$.

**Pressure dependence of $\tau_\alpha$.** In Figs. 5 - 7 the logarithm of $\tau_\alpha$ measured at various temperatures for the three samples are plotted *vs.* pressure. In all cases the behavior is linear ("volume activated") at low pressures, with the pressure sensitivity increasing for larger $\tau_\alpha$. For lower pressures

$$\tau_\alpha(P) = \tau_0 \exp\left(P\Delta V^{\#}/RT\right) \qquad (9)$$

where $\Delta V^{\#} = \ln(10) RT\, \partial \log(\tau)/\partial P\big|_T$ is the activation volume and $R$ the gas constant. Alternatively, the activation volume can be expressed as

$$\Delta V = -RT\kappa_T \frac{\partial \ln(\tau_\alpha)}{\partial \ln(V)}\bigg|_T \qquad (10)$$

where $\kappa_T$ ($=-\frac{1}{V}\frac{\partial V}{\partial P}\big|_T$) is the isothermal compressibility. Eq.(10) shows directly the relationship between $\tau$ and $V$. If the scaling (eq.(1)) is valid, it then follows that

$$\frac{\partial \ln(\tau_\alpha)}{\partial \ln(V)}\bigg|_T = \frac{\partial \ln(\tau_\alpha)}{\partial T}\bigg|_V T\gamma = -\frac{\gamma}{T}\frac{\partial \ln(\tau_\alpha)}{\partial(1/T)}\bigg|_V = -\gamma m_V(T) \qquad (11)$$

where $m_V(T)$ is the $T_g$-normalized temperature dependence at fixed $V$, which for $T = T_g$ is the isochoric fragility. Combining eqs.(10) and (11), the activation volume is related to the scaling exponent as

$$\Delta V = RT\kappa_T m_V \gamma \quad (12)$$

Since for a given material at fixed $\tau_\alpha$ (for example $\tau_\alpha(T_g)$) $m_V$ is a constant,[32] it follows that the $T$ dependence of the activation volume at fixed $\tau_\alpha$ is determined by the product $\kappa_T T$, or equivalently that $\partial \log(\tau)/\partial P|_T \propto k_T$.

If $\log(\tau_\alpha)$ varies non-linearly with pressure, an equation similar to eq.(4) with $T$ replaced by inverse pressure has been found to accurately describe the pressure dependence.[18,69] Using the analog of eq.(5) for the high pressure data [46,70]

$$\phi_P = \left\{ d\left[\log(\tau)\right]/dP \right\}^{-\frac{1}{2}} \quad (13)$$

for volume-activated behavior $\phi_P$ is a constant but otherwise varies linearly $P$. In Figures 5b and 6b is shown the function $\phi_P$ calculated respectively for the trimer and tetramer. In both cases a dynamic crossover is observed, reflected in the change in behavior of $\phi_P$. For both materials this crossover occurs at a constant value of the relaxation time, $\tau_c \sim 10^{-4}$ s independent of pressure. Interestingly, this $\tau_c$ is much larger than the relaxation time, $\tau_B$, at the crossover in the temperature dependence of $\tau_\alpha$ (Fig. 2). Previously dielectric measurements on phenolphthalein-dimethylether PDE and polychlorinated biphenyls [46,71,72] and viscosity measurements on ortho-terphenyl and salol [70] indicated that the change in slope in $\phi_P$ occurs at the same $\tau_\alpha$ (or viscosity) as the crossover occurring in $\phi_T$. This correspondence between the crossover in $\phi_P$ and $\phi_T$ is also evident in the scaling properties (eq.(1)) for various glass-forming liquids.[32,73] In the present case, however, the high pressure measurements are limited to frequencies below $10^6$ Hz, so that the existence of a second crossover in $\phi_P$ at the shorter $\tau_B$ cannot be tested. Supporting this supposition are previous cases [32,46] in which the pressure dependence at the crossover changed from VFTH-like at low pressure to approximately activated behavior at high pressure; the PMMA data in Figs. 5b and 6b show the opposite behavior. By far the most common behavior is for log $\tau_\alpha(P)$ to be linearly proportional to $P$ at low $P$.
[74,75,76,77,78,79,80,81,82,83,84,85,86,87,88,89,90,91]

For molecular liquids, in which structural relaxation involves rotation of the molecule, the activation volume, reflecting the unoccupied space necessary for the motion, often has a value close to the molar volume. For polymers, in which structural relaxation corresponds to correlated conformational transitions of several backbone bonds, there is no obvious unit to identify with the local segmental dynamics. Generally for polymers activation volumes are found to be significantly larger than the volume of the repeat unit.[76,84] Since the PMMA oligomers have the same chemical structure, their respective $\Delta V^{\#}$ should be equivalent. The insert to figure 7 shows $\Delta V^{\#}$ in the limit of zero pressure as a function of temperature normalized by the ambient pressure $T_g$. The values for the three oligomers are quite close.

*Local segmental relaxation dispersion.* Another important characteristic of structural relaxation is the shape of the dispersion (relaxation function). An equation often used to describe the dispersion is the one-sided Fourier transform of the Kohlrausch function [92,93]

$$\phi_{KWW}(t) = \exp\left[-(t/\tau_\alpha)^{\beta_{KWW}}\right] \qquad (14)$$

where $\beta_{KWW}$ is the stretching parameter ($0 < \beta_{KWW} \leq 1$). There is a general correlation between $\beta_{KWW}$ and $m$ for atmospheric pressure,[2] with larger values of fragility associated with broader dispersions. Although within a given class of glass-formers the correlation is at least approximately valid, exceptions abound more generally.[94] For non-associated materials at elevated pressure, $m$ usually decreases,[32] while $\beta_{KWW}$ is unchanged for conditions of $T$ and $P$ such that $\tau_\alpha$ is constant.[95,96]

In figure 8 we show spectra for the trimer and tetramer measured at different $T$ and $P$ chosen such that the $\tau_\alpha \sim 10$s. Good superpositioning of the peaks is found, in accord with the behavior of other glass-formers, both molecular and polymeric.[95,96] The fit of eq.(14) (shown as solid lines in the figure 8) yields $\beta_{KWW}$=0.52 for the tetramer and $\beta_{KWW}$=0.55 for the trimer. Thus, the more fragile material has a smaller $\beta_{KWW}$, in accord with the general pattern.[2] For the decamer the intensity of the secondary relaxation is relatively large, so that a distinct loss peak is never isolated away from the secondary relaxation. Since the $T$ and $P$ dependences of the relaxation times and intensities for the two peaks may differ, their deconvolution is necessary to accurately determine the shape of the local segmental peak. However, the proper method to deconvolute is controversial.[39,97,98,99]

*Equation of state.* Above the glass transition experimental specific volumes can be represented using the Tait equation of state (EOS)

$$V(T,P) = \left(a_0 + a_1 T + a_2 T^2\right)\left\{1 - 0.0894 \ln\left[1 + \frac{P}{b_1 \exp(-b_2 T)}\right]\right\} \quad (15)$$

where $a_0$, $a_1$, $a_2$, $b_1$ and $b_2$ are constants having the values listed in Table 2. (By convention temperature is in units of Celsius.) Using the respective EOS $\tau_\alpha$ for each sample is obtained as a function of $V$ (figure 9). As found for virtually all glass-forming materials,[18] the relaxation times are not defined solely by $V$. However, the data do superimpose when plotted versus the $TV^\gamma$ (Figure 10). Also included in Fig. 10 are $\tau_\alpha$ for a high molecular weight PMMA ($M_w=1.5\times10^5$g/mol) measured by Theobald et al.[61] We find that the scaling exponent decreases with increasing molecular weight: $\gamma = 3.7$, 3.2, 2.8 and 1.8 for the trimer, tetramer, decamer, and high polymer, respectively. In the case of high $M_w$ PMMA it is possible to determine the pressure dependence of $T_g$ from PVT measurements and estimate $\gamma$. The ratio $E_V/H_P$ can be calculated using[76]

$$\frac{E_V}{H_P} = \frac{1}{1 - \alpha_P/\alpha_\tau} \quad (16)$$

where $\alpha_\tau \left(= \partial \ln(V)/\partial T\big|_\tau\right)$ is the isochronic thermal expansion coefficient. For high Mw PMMA we find $\alpha_\tau = (-1.4\pm0.2)\times10^{-3}$ C$^{-1}$, from which we calculated $E_V/H_P = 0.73\pm0.03$. From eq.(3) we then calculate $\gamma = -\left(\alpha_\tau T_g\right)^{-1} = 1.85\pm0.25$ that is in good agreement with $\gamma$ found by superpositioning the $\tau$ data. In the literature we found a somewhat lower value of $\gamma = 1.25$ was reported for a different high $M_w$ PMMA [100]) but no details were provided concerning the molecular weight or tacticity of the sample used in that study.

For the trimer $\gamma$ is close to the value for van der Waals molecular glass-formers such as orthoterphenyl ($\gamma = 4$),[101] propylene carbonate ($\gamma=3.7$),[102] cresolphthalein-dimethylether ($\gamma=4.5$),[73] phenylphthalein-dimethylether ($\gamma=4.5$),[20] and decahydroisoquinoline ($\gamma=3.55$),[103] although values of $\gamma$ as high as 8.5 have been found for bulkier molecular liquids, such as 1,1'-di(4-methoxy-5-methylphenyl)cyclohexane [20] and polychlorinated biphenyls.[90]

**Discussion**

This study of local segmental relaxation in PMMA makes clear that chain length exerts a strong influence on properties, in particular the relative effect of $T$ and $V$ on the dynamics. In going from the trimer to the high polymer, the effect of volume is suppressed, as reflected in the decrease of $\gamma$ from 3.7 to 1.8. Qualitatively we interpret this as due to a softening of the potential due to an increasing influence of the intra-chain potential. There is a concomitant reduction in the packing efficiency with $M_w$, as seen in the increasing volume at the glass transition, $V_g$, with $M_w$ (table 3). These chain length dependences are consistent with the relative stiffness of the PMMA chain.

In Table 3 are listed the activation volumes for three oligomers and the high molecular weight PMMA polymer as obtained using eq.(12). The larger $\Delta V^{\#}$ for the high polymer follows from its higher $T_g$ and larger $\kappa_T$. The molar volume of the repeat unit in PMMA determined at $T_g$ using the equation of state (described below with parameters in Table 3) are 78, 80 and 83 ml/mol for $n=3$, 4 and 10, respectively, which is about 50% smaller than $\Delta V^{\#}$. Another obvious difference in comparing $\Delta V^{\#}$ with actual volume $V$, is that $\Delta V^{\#}$ increase with decreasing $T$ while $V$ decreases, so that in the limit of high $T$ (~1.3$T_g$) the two have comparable magnitude. This is consistent with results for other polymers.[16,104,105]

Previously we showed for propylene carbonate, salol, polyvinylacetate, o-terphenyl and a mixture of o-terphenyl with o-phenyl phenol that the entropy is well represented by [23,24]

$$S = f(TV^{\gamma_S}) \qquad (17)$$

with the scaling exponent $\gamma_S$ bears a relationship to the Grüneisen parameter, $\gamma_G$, defined as

$$\gamma_G = \frac{V\alpha_P}{C_V\kappa_T} \qquad (18)$$

The condition $TV^{\gamma_S} = constant$ corresponds to a reversible adiabatic transformation reminiscent of an ideal gas. Interestingly, the scaling exponent for the relaxation times, $\gamma_\tau$, is about threefold larger than $\gamma_G$,[22,23, 24] due to non-configurational contributions to the entropy $S_0$, from vibrational and local secondary motions, which do not affect structural relaxation.

It is reasonable to assume that in the liquid state during an isothermal volume change, the entropy change is purely configurational; that is, the unoccupied or "free" volume has to be removed before vibrational or local intramolecular motions are affected. A simple parallel is that of a soft matrix containing hard particles - compressing the matrix is not expected to appreciably change volume of the particles. In this approximation then

$$\left.\frac{\partial S_c}{\partial V}\right|_T = \left.\frac{\partial S_{liq}}{\partial V}\right|_T - \left.\frac{\partial S_0}{\partial V}\right|_T \approx \left.\frac{\partial S_{liq}}{\partial V}\right|_T \qquad (19)$$

where $S_{liq}$ is the total entropy of the liquid, $S_c$ the entropy in excess of that from vibrational or local intramolecular motions, and $S_0$ the non-configurational entropy. $S_c$ is calculated starting from the differential form

$$dS_c = \left(\left.\frac{\partial S_{liq}}{\partial T}\right|_V - \left.\frac{\partial S_0}{\partial T}\right|_V\right)dT + \left(\left.\frac{\partial S_{liq}}{\partial V}\right|_T\right)dV = \frac{\Delta C_V}{T}dT + \left.\frac{\partial P_{liq}}{\partial T}\right|_V dV \qquad (20)$$

where we have used eq.(19) together with one of the Maxwell relations, and $\Delta C_V = C_V^{liq} - C_V^{non}$. Defining the Grüneisen parameter for $S_c$, $\gamma_{Sc}$ as

$$\gamma_{Sc} = \frac{V\alpha_P^{liq}}{\Delta C_V \kappa_T^{liq}} \qquad (21)$$

equation (20) can be rewritten as

$$dS_c = \Delta C_V\left(\frac{dT}{T} + \gamma_{Sc}\frac{dV}{V}\right) \qquad (22)$$

If $\gamma_{Sc}$ and $\Delta C_V$ are respectively independent of $V$ and $T$, integration of eq.(22) gives

$$S_c = \Delta C_V \ln\left(TV^{\gamma_{Sc}}\right) + const \qquad (23)$$

This implies that $S_c$ scales as a function of the variable $TV^{\gamma_{Sc}}$, in the manner to $\tau_\alpha$, which in turn implies an entropy basis for the glass transition dynamics.

In Table 3 are the values of $\gamma_{Sc}$ calculated using eq.(21). Since the PMMA are all atactic, there is no crystalline state and so the available reference is the glass. This approximation underestimates $S_c$, since some configurational degrees of freedom may not be frozen in the glass (possibly manifest as secondary relaxations). Since $V(T,P)$ data are unavailable for the samples below $T_g$, we make the further approximation that $C_V^{glass} \sim C_P^{glass}$; thus, we take $\Delta C_V \cong C_P^{liq} - C_P^{glass}$. The obtained $\gamma_{Sc}$ are larger than the values of $\gamma$ that scale $\tau_\alpha$; however, both decrease in a similar fashion with increasing $M_w$, unlike $\Delta C_P$, which is approximately constant. This variation with $M_w$ could be related to the increasing contribution of the secondary relaxation to the entropy, which contributes to an underestimate of $\Delta C_V$ and $S_c$.[106] Note that the difference between $\gamma_{S_c}$ and $\gamma$ increases with $M_w$, in parallel with the increase of the relative strength of the secondary relaxation.[107]

The best known model for the glass transition is due to Adam and Gibbs,[108,109] who predicted $\tau_\alpha$ to be a function of the product $TS_c$. In typical measurements near the glass transition, $\tau_\alpha$ changes by ~ 8 orders of magnitude, while $T$ changes *ca.* 20% and $S_c$ about two orders of magnitude. This implies that the dynamics are dominated by changes of $S_c$. A variation on the model of Adam and Gibbs is due to Avramov.[110] We have shown that from eq.(23) for the entropy, the following expression for $\tau_\alpha(T,V)$ is obtained [22]

$$\log[\tau(T,V)] = A + \left(\frac{B}{TV^{\gamma_A}}\right)^D \qquad (24)$$

where A, B, D, and $\gamma_A$ are constants. This equation, shown in Fig.9 as solid lines, accurately describes the $\tau_\alpha$ over a broad dynamic range, with the obtained parameters given in Table 4. The $\gamma_A$ determined from fitting eq.(24) are essentially equivalent to the $\gamma$ obtained by superpositioning of the $\tau_\alpha$ data.

In assessing entropy models for polymers, there is a lack of connection between the magnitude of the heat capacity change at $T_g$ and the variation of fragility with $M_w$.[64,68] However, considering both the $T$ and $V$ dependences of $S_c(T,V)$, the calculated value of $\gamma_{S_c}$ is close to that of $\gamma$, and has the same trend with changing $M_w$. In fact $\gamma_{S_c}$ depends inversely on the $\Delta C_P$ and directly on the ratio $\alpha_P/\kappa_T$, and in the same manner as $\gamma$, decreasing with increasing $M_w$. The change of $\Delta C_P$ from low to high $M_w$ is only ~10%, while the change of the ratio $\alpha_P/\kappa_T$ is about one-half, similar to the change in $\gamma$.

Application of eq.(5) indicates a crossover for the trimer and, more weakly, for the tetramer, at a relaxation time $\tau_B \sim 10^7$Hz (fig.2). This crossover time was found to be independent of pressure in other materials,[46,70] but that could be investigated herein because of limited dynamic range of the high pressure measurements. However, using eq.(13) a second crossover at much lower frequency $\tau_\alpha \sim 10^{-4}$Hz is found (figs. 5-6). This pressure crossover transpires between the volume activated dynamics at short times ($\Delta V^{\#}$~const) and the VFTH-like dependence at longer times ($\Delta V^{\#}$ increases with P). This is different from the pressure crossover behavior found previously, for which there was a correspondence between the relaxation times from eqs. (5) and (13). Since the $TV^{\gamma}$ scaling (eq.(1)) applies to the trimer and tetramer, we can use the atmospheric pressure $\tau_{\alpha}(T,V)|_{P=0.1MPa}$ (which could be measured over a broad range) together with the EOS to calculate the isothermal behavior, $\tau_{\alpha}(P,V)|_{T=const}$. For a given temperature $T=T_A$, this is done first using eq.(1) to calculate for each $\tau_{\alpha}(T,V)|_{P=0.1MPa}$ the $V$ for which $\tau_{\alpha}(P,V)|_{T=T_A}$. Then the corresponding pressure $P(T_A,V)$ is obtained from the EOS (negative pressure results were not considered). As shown previously for other materials,[32] this procedure yields an accurate set of isothermal $\tau_{\alpha}$, from which the crossover could be determined. Eq.(13) is applied to these data, as shown in Fig.11 as solid symbols, together with the $\phi_P$ in Figs. 5 and 6 calculated from the actual high pressure measurements (open symbols). The consistency between these two sets of data (in their common range of $P$) shows that the behavior is intrinsic to the atmospheric pressure and EOS results. Note that for the trimer there is a hint of a second crossover at $\tau \sim \tau_B$, at least for the higher $T$ data set. The second crossover at longer $\tau_{\alpha}$ could be a signature of the splitting of the local segmental and secondary relaxations. This will be the subject of future work.[107]

**Conclusions**

The structural dynamics of PMMA oligomers ($n$ = 3, 4, and 10) was studied over a wide range of frequency as a function of $T$, $P$, and $V$. The dependence of the structural relaxation on $T$ and $V$ is strongly influenced by the chain length, as reflected in the variation of $\gamma$ with $M_w$. For the smallest polymer chain (trimer), the scaling exponent is very close to that of for van der Walls liquids ($\gamma \sim 4$), and it decreases with increasing $M_w$ to $\gamma=1.8$ for the high polymer. This shows that small values of the scaling parameter, as generally observed for polymers,[18] are mainly due to the relative stiffness of the backbone units, in comparison to the mobility arising from intermolecular degrees of freedom. We also find, consistent with previous works for a given family of materials,[111,112,113] a similar effect of $M_w$ on other dynamic properties, such as the fragility and the glass transition temperature, and on thermodynamic parameters such as the heat capacity change and the specific volume at $T_g$. The chain structure hinders segment rearrangements with a consequent smaller sensitivity to intermolecular distance and larger $V$. Conversely, polymers having a very flexible chain (e.g., siloxane polymers) are associated with larger values of γ (~5) and near invariance of $m$ and $T_g$ to $M_w$.

Using the derivation function (eq.(5)), we observed a crossover in the temperature dependence of the relaxation times at atmospheric pressure for the trimer and tetramer at $\tau_B$~$10^{-7}$s. However for the latter, this crossover is weaker, suggesting that for sufficiently high $M_w$ it may eventually "disappear". Unfortunately for high $M_w$ PMMA the substantial intensity the secondary relaxation precludes testing this supposition. Using eq.(13) we observe both the crossover around $\tau_\alpha \sim \tau_B$, and a second crossover at much longer $\tau_\alpha$ (~$10^{-4}$s). The latter does not seem to have a counterpart in the $T$ behavior of $\phi_T$. However, at this same value of the relaxation time there is a decoupling between the conductivity and the local segmental relaxation (insert to Fig.3).

Finally, for the PMMA oligomers we find that the shape of the relaxation dispersion, as described using a KWW function, is constant for given value of the relaxation time; that is, the segmental dispersion is invariant to different thermodynamical conditions at constant $\tau_\alpha$. Together with the scaling (eq.(1)), this means that γ determines both the relaxation time and the breadth of its distribution.

**Acknowledgments**

This work was supported by the Office of Naval Research and by MIUR (PRIN 2005). The authors thank K.J. McGrath for providing PVT data and L.J. Buckley for use of the HP4291A impedance analyzer.

**Figure captions**

Figure 1. (a) Selected dielectric loss spectra for the trimer measured at atmospheric pressure for temperatures (from left to right): 211, 215, 218, 223, 228, 233, 238, 243, 248, 253, 262 K. (b) Selected dielectric loss spectra for the tetramer measured at atmospheric pressure for temperatures (from left to right): 246, 253, 263, 273, 283, 293, 313, 333, 358 K.

Figure 2. (a) $\alpha$-relaxation time for PMMA oligomers having the indicated number of repeat units. The solid lines represent the VFTH equation, with the parameters given in Table 1. The inset shows the deviation between the VFTH and the experimental data for the tetramer. (b) Stickel derivative plots. The arrows indicate the dynamic crossover, corresponding to intersection of the fitted lines for the trimer and the deviation from the VFTH for the tetramer.

Figure 3: DC conductivity *vs.* reciprocal temperature for PMMA having the indicated degree of polymerization. In the inset is plotted the conductivity *vs.* segmental relaxation time; the solid line indicates a slope of minus unity.

Figure 4: Molecular weight dependence of atmosphere pressure $T_g$ measured herein by calorimetry and dielectric relaxation ($\tau(T_g)$=100s), and from DSC measurements of Faetti et al.[57]. The solid and dotted lines are the respective best fits to the UK ($T_{g,\infty}$=393±4 K, $k_{UK}$=0.7±0.04 g mol$^{-1}$K$^{-1}$) and FF ($T_{g,\infty}$=381±8 K, $k_{FF}$=(5.6±0.4)×10$^4$ g mol$^{-1}$K), equations. The inset shows the change with molecular weight of the heat capacity at the glass transition from measurements herein and literature data,[57,61,62,63]; the dotted line is to guide the eyes.

Figure 5: (a) Isothermal segmental relaxation times *vs.* pressure for the trimer. (b) derivative function for the pressure dependence; solid lines are a guide to the eyes.

Figure 6: (a) Isothermal segmental relaxation time *vs.* pressure for the tetramer. (b) derivative function for the pressure dependence; solid lines are a guide to the eyes.

Figure 7: Isothermal segmental relaxation time *vs.* pressure for the decamer. The insert shows the activation volume in the limit of zero pressure as a function of temperature normalized by the glass temperature (at $P$ = 0.1 MPa). $\Delta V$ was determined from linear fits of the isothermal log($\tau_\alpha$) *vs.* pressure data for the three oligomers; solid line is to guide the eyes.

Figure 8: Comparison of isochronal spectra ($\tau\sim$10s) for the trimer and tetramer.

Figure 9: Isothermal and isobaric segmental relaxation times *vs.* volume for the three oligomers. The solid lines represent the fit to eq.(24).

Figure 10: Scaling plots of the segmental relaxation times for the three oligomers and a high molecular weight PMMA (the latter from Theobald et al.[61]).

Figure 11. Derivative function for the pressure dependence of the trimer and tetramer. The open symbols were calculated from the isothermal measurements (as in Figs. 5 and 6), while the solid symbols were calculated from the atmospheric data using the scaling properties (see text). Solid lines are to guide the eyes.

| $n$ | $\log(\tau_0[s])$ | $T_0[K]$ | $D$ | $m$ |
|---|---|---|---|---|
| 3  ($T<T_B=264K$) | $-14.5 \pm 0.2$ | $161.6 \pm 1$ | $11.0 \pm 0.4$ | $74 \pm 2$ |
| 3  ($T>T_B=264K$) | $-12.1 \pm 0.1$ | $196.1 \pm 2.5$ | $4.1 \pm 0.3$ | |
| 4 | $-13.+05 \pm 0.03$ | $195.3 \pm 0.3$ | $7.9 \pm 0.1$ | $81 \pm 2$ |
| 10 | $-10.9 \pm 0.2$ | $238 \pm 4$ | $6.1 \pm 0.5$ | $75 \pm 10$ |
| $M_w=1.5\times10^{5\ b}$ | $-11 \pm 1.9$ | $337 \pm 14$ | $3.8 \pm 1.9$ | $115 \pm 16$ |

[a] $\tau(T_g) = 100$ s     [b] From data of Theobald et al. [61].

Table 1. Fit parameters for eq. (4).

| $n$ | $a_0$(ml/g) | $A_1$(ml/gC) | $a_3$(ml/gC$^2$) | $b_1$(MPa) | $b_2$(C$^{-1}$) |
|---|---|---|---|---|---|
| 3 | $0.817\pm0.0002$ | $(6.43\pm0.07)\times10^{-4}$ | $(5.0\pm0.6)\times10^{-7}$ | $185\pm0.6$ | $(5.2\pm0.05)\times10^{-3}$ |
| 4 | $0.819\pm0.0002$ | $(6.33\pm0.06)\times10^{-4}$ | $(5.2\pm0.6)\times10^{-7}$ | $191\pm0.6$ | $(4.7\pm0.05)\times10^{-3}$ |
| 10 | $0.818\pm0.0001$ | $(6.05\pm0.01)\times10^{-4}$ | - | $235\pm0.4$ | $(4.3\pm0.02)\times10^{-3}$ |
| $1.5\times10^{3\ a}$ | $0.819\pm0.002$ | $(3.2\pm0.3)\times10^{-4}$ | $(6.1\pm1)\times10^{-7}$ | $316\pm8$ | $(4.6\pm0.1)\times10^{-3}$ |

[a] From analysis of PVT data in ref. [61]

Table 2. Fit parameters for eq.(15).

| $n$ | $T_g$ (K) | $\alpha_P\times10^{-4}$ (K$^{-1}$) | $\kappa_T\times10^{-4}$ (MPa$^{-1}$) | $V_g$ (mlg$^{-1}$) | $\Delta V_g$ (mlmol$^{-1}$) | $\Delta V_g/kT_g$ (kJmol$^{-1}$K$^{-1}$) | $\gamma$ | $E_V/E_P\vert_{Tg}$ | $\Delta C_P\vert_{Tg}$ (JK$^{-1}$g$^{-1}$) | $\gamma_{Sc}$ |
|---|---|---|---|---|---|---|---|---|---|---|
| 3 | 209.6 | 7.48 | 3.47 | 0.774 | 132±9 | 1.8±0.1 | 3.7 | 0.63 | 0.36 | 4.6 |
| 4 | 239.6 | 7.50 | 4.00 | 0.797 | 130±3 | 1.4±0.1 | 3.2 | 0.63 | 0.36 | 4.1 |
| 10 | 287.9 | 7.32 | 4.05 | 0.827 | 128±17 | 1.1±0.1 | 2.8 | 0.63 | 0.36 | 4.1 |
| $1.5\times10^3$ | 379.8 | 5.24 | 4.62 | 0.860 | 216±30 | 1.2±0.2 | 1.8 | 0.74 | 0.29 | 3.3 |

Table 3. Thermodynamic parameters for the three oligomers and a high $M_w$ PMMA ($n=1.5\times10^3$): glass transition temperature, $T_g$; isobaric expansion coefficient, $\alpha_P$; isothermal compressibility, $\kappa_T$; specific volume at $T_g$, $V_g$; activation volume at $T_g$, $\Delta V_g$; scaling parameter of the relaxation time, $\gamma$, ratio of the isochoric and isobaric activation energies (eq.(3)); change of the isobaric heat capacity at the glass transition; scaling parameter of $S_c$ estimated as $\gamma_{sc}=V\alpha_P/\kappa_T\Delta C_P$.

| $n$ | $A$ | $B$ | $D$ | $\gamma_A$ |
|---|---|---|---|---|
| 3 | $-9.53\pm0.08$ | $83.3\pm0.2$ | $3.91\pm0.06$ | $3.60\pm0.01$ |
| 4 | $-9.30\pm0.07$ | $116.2\pm0.4$ | $4.07\pm0.06$ | $3.22\pm0.01$ |
| 10 | $-8.5\pm0.2$ | $163.4\pm1.2$ | $3.1\pm0.2$ | $2.98\pm0.04$ |

Table 4. Fit parameters for eq.(24) (shown as solid lines in Fig. 9).

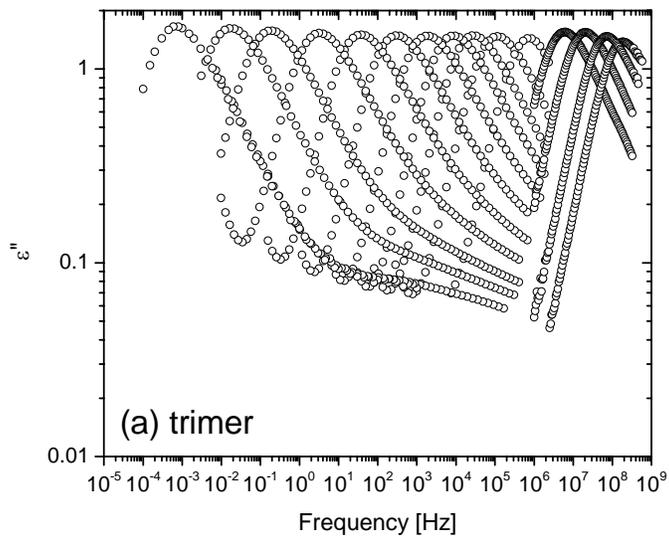

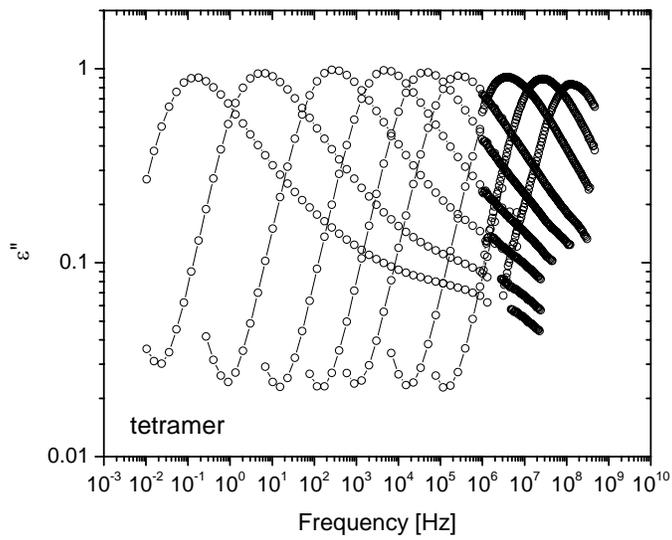

Figure 1

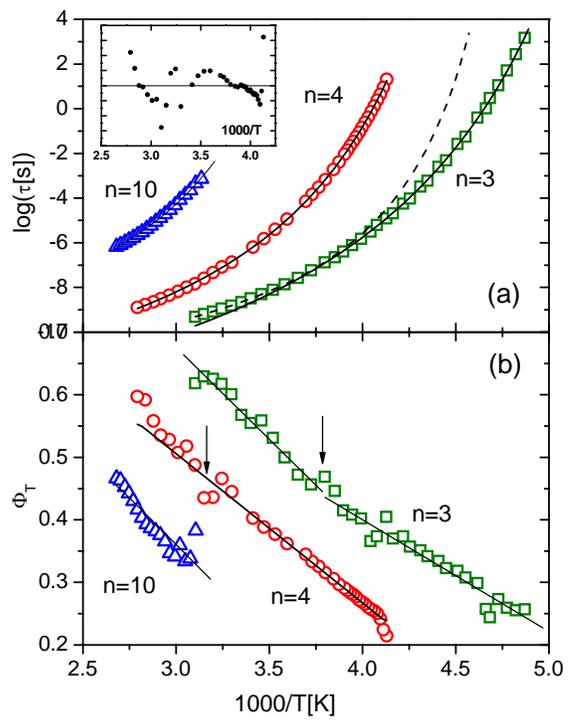

Figure 2

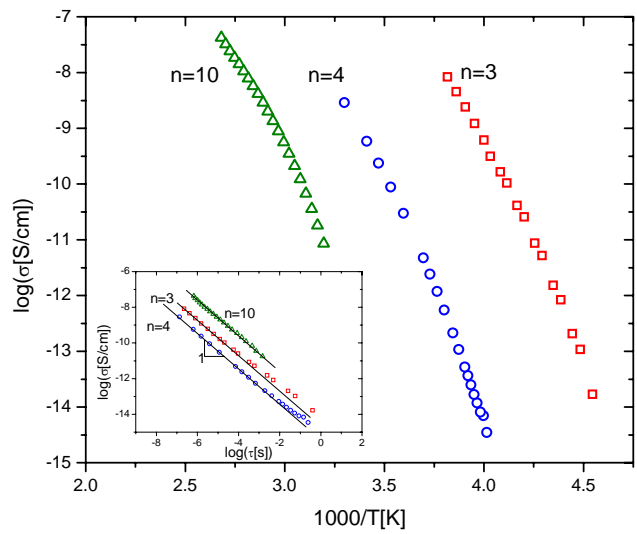

Figure 3

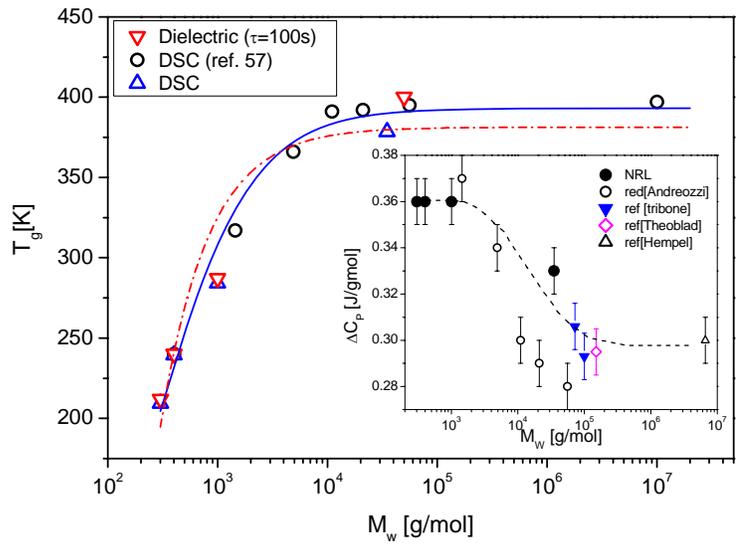

Figure 4

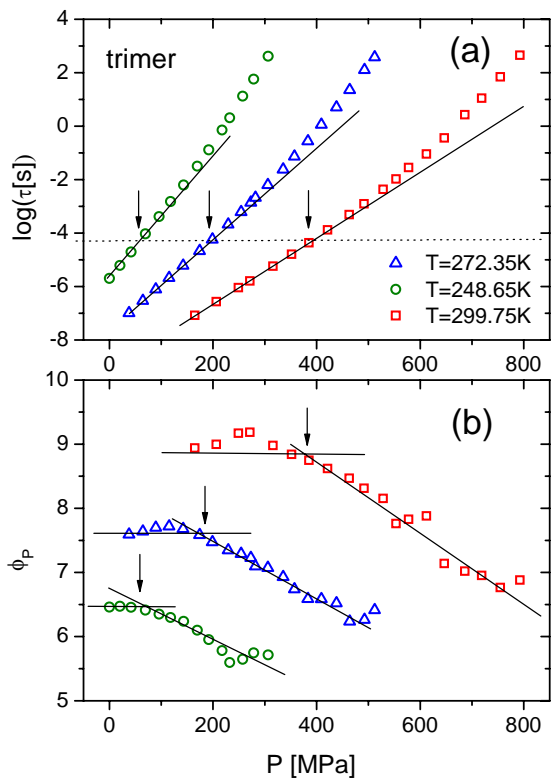

Figure 5

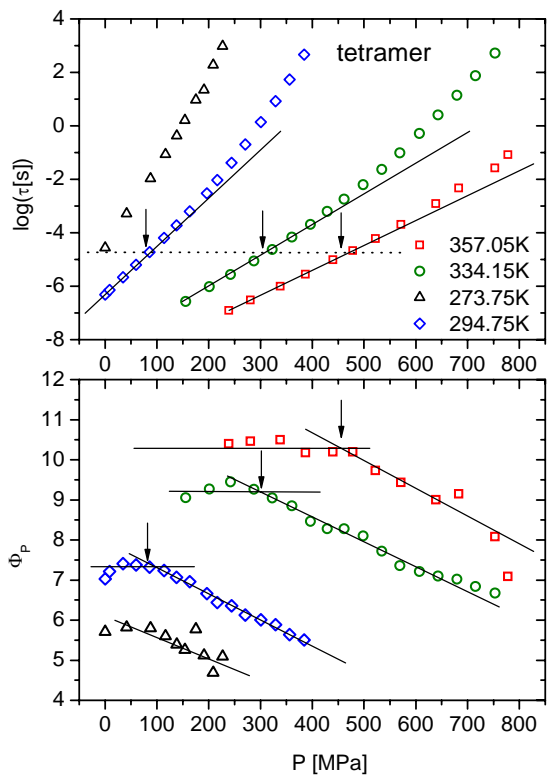

/

Figure 6

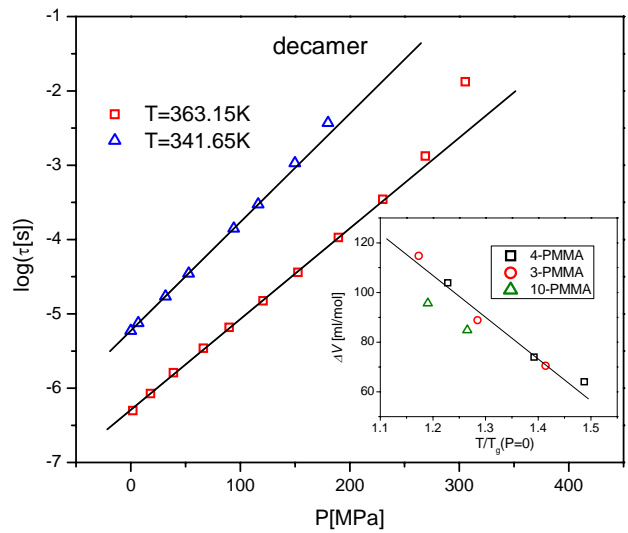

Figure 7

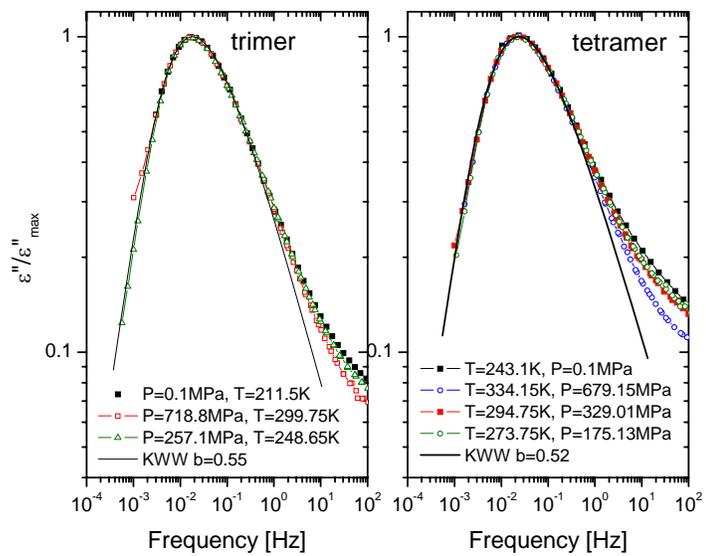

Figure 8

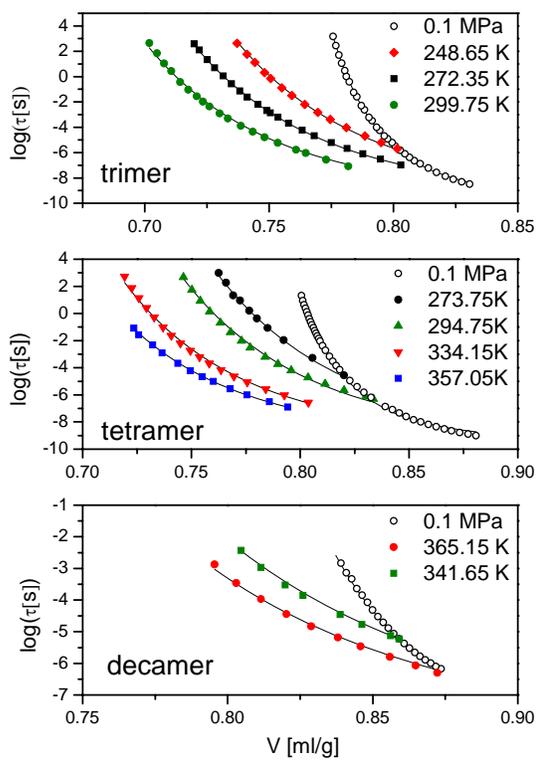

Figure 9

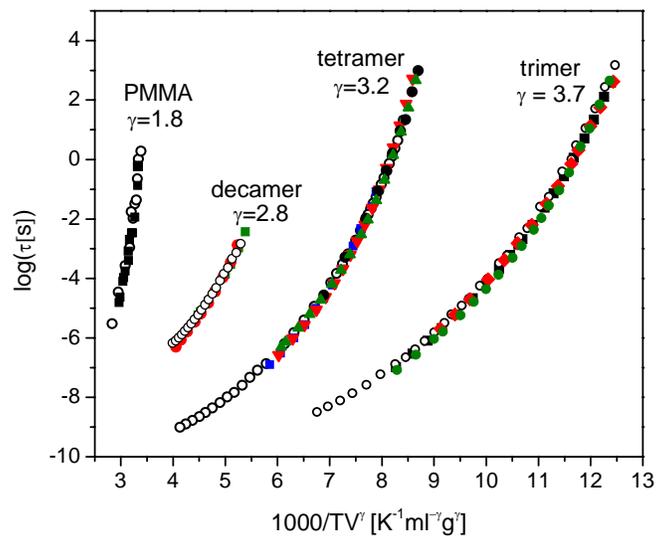

Figure 10

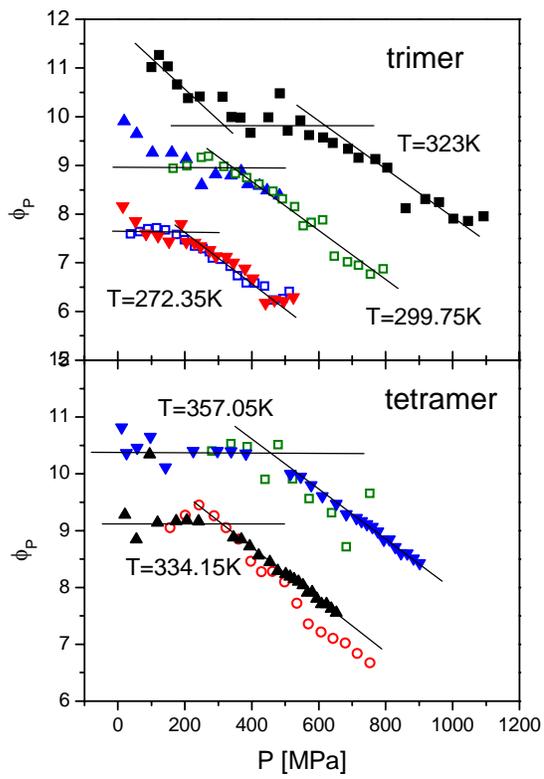

Figure 11